# An Augmented Load–Frequency Control Block Diagram for Power Systems

Mahyar Tofighi-Milani, Sajjad Fattaheian-Dehkordi, and Matti Lehtonen

***Abstract*-- The increasing penetration of inverter-based resources has led to a significant reduction in system inertia, resulting in faster and more pronounced frequency deviations in modern low-inertia power systems. In such environments, the dynamic behavior of electrical loads becomes increasingly important in shaping overall system frequency response. This paper presents an enhanced load model that incorporates load-side dynamics in addition to conventional static behavior, thereby augmenting the representation of load–frequency control (LFC) models. This improved formulation increases the accuracy of frequency response studies in power systems. A comparison between the proposed augmented model and the conventional LFC representation demonstrates that relying solely on static load modeling can lead to inaccurate results and potentially misleading conclusions. Therefore, accurate modeling of load-side dynamics is essential for reliable frequency stability assessment in modern power systems.**

## I. Introduction

In future power systems dominated by inverter-based resources, the reduced presence of synchronous generators lowers system inertia, leading to faster and larger frequency deviations following disturbances. As a result, the impact of electrical loads on frequency response becomes increasingly significant. Accurate load modeling is therefore essential to properly capture their contribution and ensure reliable frequency stability [1].

Electrical loads in power systems can be broadly classified into frequency-independent loads and frequency-dependent loads (FDLs). While frequency-independent loads maintain constant power consumption regardless of frequency variations, FDLs exhibit power consumption that varies with system frequency [2]. Moreover, FDLs can be categorized into static and dynamic components. Static FDLs are typically represented through algebraic relationships with frequency, whereas dynamic FDLs depend on both frequency and time, reflecting their transient behavior [3], [4].

Despite this classification of loads, conventional load representations remain widely used in load–frequency control (LFC) block diagrams within power system dynamics studies. Numerous works continue to rely on this simplified framework to evaluate system performance under low-inertia conditions. For instance, in [5], [6], and [7], the frequency regulation capability of emerging flexible resources is assessed within a LFC block diagram with conventional load model. Similarly, [8] develops a power system model incorporating large-scale controllable loads to evaluate their regulation potential, while still adopting the simple damping factor in the LFC structure for frequency response analysis. The impact of distributed energy resources and microgrids on the LFC problem is investigated in [9], where their contribution is analysed within a LFC block diagram with the simple damping factor.

Furthermore, converter-based virtual inertia and grid-forming control strategies are proposed in [10] and [11] to enhance frequency nadir and rate of change of frequency (RoCoF). Also, synthetic inertia procurement is studied in [12] using standard frequency-response metrics within a conventional system-level frequency model.

At the operational planning level, frequency-response models with simplified load model are also widely adopted for frequency-secure scheduling and commitment problems. For example, [13] formulates a stochastic unit commitment framework that co-optimizes energy, reserves, and inertia-dependent fast frequency response by embedding dynamic frequency constraints associated with RoCoF, frequency nadir, and quasi-steady-state frequency. Similarly, [14] derives an analytical frequency nadir formulation to define a frequency security margin and incorporates its linearized representation into a frequency-constrained unit commitment model. Despite their effectiveness for scheduling applications, both studies rely on reduced-order system frequency models with simplified load damping representations.

In addition, several references work on LFC performance through advanced control design while still relying on the simple damping factor in the LFC load model. For example, [15] and [16] propose dynamic event-triggered control strategies for multi-area power systems under cyber-attacks, and [17] investigates time-delay and sampling effects in LFC design.

Collectively, these studies highlight the increasing importance of fast-responsive and flexible resources in enhancing frequency stability in low-inertia power systems, while still largely relying on simple damping factor representations for system modeling and analysis. In these LFC models, FDLs are typically approximated by a single damping factor, which fails to capture the detailed behavior of both static and dynamic load components. In fact, the coefficient of damping factor represents the static load of the system considered as the ZIP load model fails to model the dynamic loads. Thus, dynamic loads, i.e., induction motors—representing a significant share of real-world demand—are frequently neglected, despite their substantial influence on system frequency dynamics.

Consequently, relying solely on simplified load models may lead to inaccurate characterization of load contributions and potentially misleading conclusions in frequency response analysis. This limitation underscores the need for more accurate load-side modeling to ensure reliable frequency stability assessment in modern power systems.

In this regard, this paper proposes an augmented LFC block diagram by enhancing the load component of the conventional LFC model. In addition to the static load representation, the proposed approach incorporates the dynamic behavior of induction motors, thereby capturing both static and dynamic load characteristics. This enhancement results in a more accurate LFC representation and improves the fidelity of frequency response analysis in power systems.

## II. System structure and load modelling

This section presents an enhanced load model to augment the conventional LFC framework for dynamic power system studies. It includes an overview of the conventional system representation, the load modeling, and the overall system block diagram presentation.

### *A. Conventional System Block Diagram*

The conventional block diagram model of a power system, which is commonly used for LFC studies, is shown in Fig. 1. Where, $R$ and $T_G$ show the governor characteristics, and $F_{HP}$, $T_{RH}$ , as well as $T_{CH}$ indicate the parameters of turbine. $H$ is the system inertia, and $D$ demonstrates the damping factor. Finaly, $\overline{\Delta f}$ , $\overline{\Delta P_{FDL}}$ , and $\overline{\Delta P_L}$ are representative of frequency deviation, FDL power deviation, and load power deviation in per-unit, respectively.

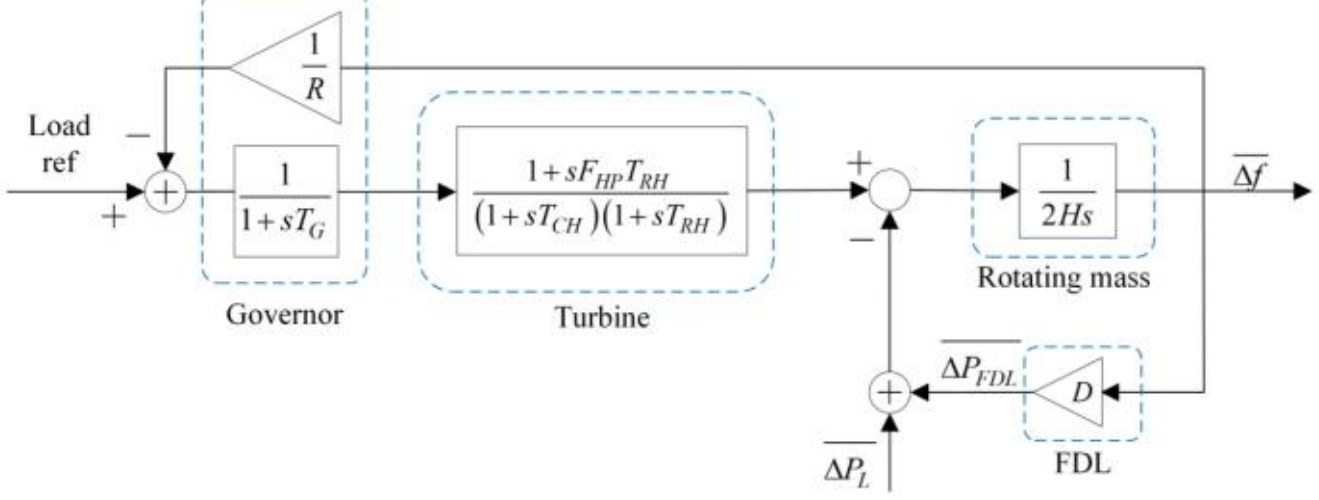

Fig. 1. The conventional block diagram of the power system

In the model represented in Fig. 1, the load sector of the power system is considered as consisted of only static loads and is considered with a constant damping factor (i.e., D), which is derived from assigning a ZIP model to the FDL sector [18]. However, not all power system loads conform to the behavior described by static load models, as their response may depend on time as well as frequency itself.

### *B. Load Modelling*

In this section, the detailed mathematical modeling of FDLs in power systems is presented. First, static load modeling is discussed, followed by dynamic load modeling. Finally, the overall load model is constructed by combining these components.

#### *1) Static Load Modeling*

As shown in the conventional block diagram in Fig. 1, the static load can be modeled via ZIP model, which responds rapidly to changes in the system frequency. The mathematical formulation of the ZIP model is represented as follows [3], [19], [20].

$$P_{ZIP} = P_{ZIP,n}\left(1 + k_{ZIP}\overline{\Delta f}\right) \quad (1)$$

Where, $P_{ZIP}$ and $P_{ZIP,n}$ demonstrate the actual and nominal Power consumption by ZIP load model, respectively. Additionally, $k_{ZIP}$ is the parameter of ZIP load, and $\overline{\Delta f}$ shows the frequency deviation in per-unit.
Equation (1) could be also represented in Laplace domain as in (2), defining $\overline{\Delta P_{ZIP}}$ as the ZIP load power deviation in per-unit.

$$\overline{\Delta P_{ZIP}}(s) = k_{ZIP}\overline{\Delta f}(s) \quad (2)$$

Based on (2), the transfer function of the ZIP load can be expressed as $H_{ZIP}$ as in (3).

$$H_{ZIP}(s) = \frac{\overline{\Delta P_{ZIP}}(s)}{\overline{\Delta f}(s)} = k_{ZIP} \quad (3)$$

From (2) and (3), it is evident that if the load of the entire system is considered as a ZIP model, the system load can be modeled by the damping factor (i.e., $k_{ZIP} = D$) as depicted in the conventional system shown in Fig. 1.

#### *2) Dynamic load modeling*

Induction motors (IMs), which represent a substantial share of industrial loads, are the most predominated dynamic loads in power systems. In the Laplace domain, the relationship between frequency variations and the power consumption of IMs (i.e. $\overline{\Delta P_{IM}}$ ) could be described by the following equation [21]:

$$\overline{\Delta P_{IM}}(s) = \frac{k_{IM,n} + k_{IM,r}\cdot s}{1+\tau_{IM}\cdot s}\overline{\Delta f}(s) \quad (4)$$

Where, $k_{IM,n}, k_{IM,r}, \tau_{IM}$ are the parameters of the IM model.

Equation (4) could be also shown as (5) in terms of IM transfer function (i.e. $H_{IM}$ ).

$$H_{IM}(s) = \frac{\overline{\Delta P_{IM}}(s)}{\overline{\Delta f}(s)} = \frac{k_{IM,n} + k_{IM,r}\cdot s}{1+\tau_{IM}\cdot s} \quad (5)$$

#### *3) Block Diagram of the total load model*

By integrating the IM model into load-sector modeling of the power system, dynamic load behavior can be incorporated into the LFC framework. Considering both static and dynamic load models enables a more accurate representation of FDLs in LFC.

To account for both load components, the parameter $\alpha_{IM}$ is defined as the proportion of IM load within the total FDL demand. Accordingly, the system FDL can be reformulated as given in (6), or equivalently expressed in transfer-function form as in (7).

$$\overline{\Delta P_{FDL}}(s) = \left[\left(1-\alpha_{IM}\right)H_{ZIP}(s) + \alpha_{IM}H_{IM}(s)\right]\overline{\Delta f}(s) \quad (6)$$

$$D(s) = \frac{\overline{\Delta P_{FDL}}(s)}{\overline{\Delta f}(s)} = \left(1-\alpha_{IM}\right)H_{ZIP}(s) + \alpha_{IM}H_{IM}(s) \quad (7)$$

Where, $\overline{\Delta P_{FDL}}$ denotes the power deviation of the FDL, while $D(s)$ represents the transfer function of the load sector. It should be noted that, in conventional system modeling, the load damping term is typically represented by the constant $D$.

However, with the inclusion of dynamic load effects, this term is generalized to the enhanced damping function $D(s)$. This implies that, for greater modeling accuracy, it is more appropriate to use $D(s)$ instead of the conventional constant $D$. The proposed enhanced FDL model for the power system is also illustrated in the block-diagram schematic shown in Fig. 2.

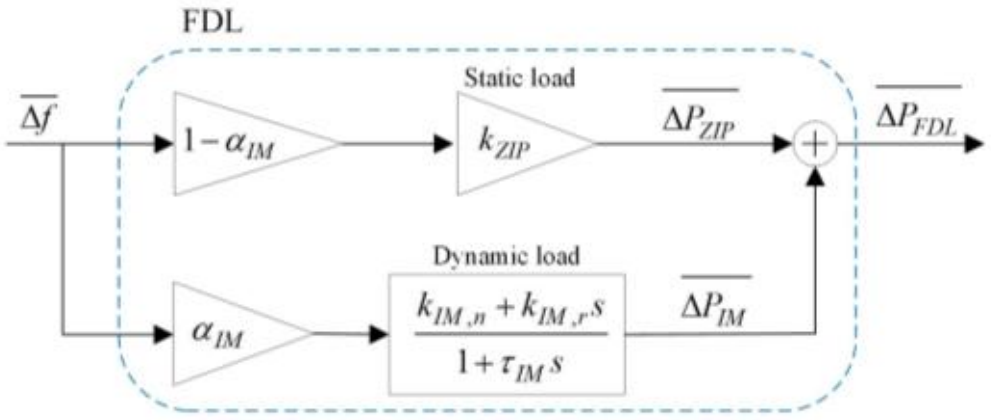

Fig. 2. The enhanced FDL model for power systems.

By the implementation of the enhanced FDL model in the LFC model, one could represent the overall system block diagram as shown in Fig. 3. Where, $D(s)$ is presented through the transfer function in (7).

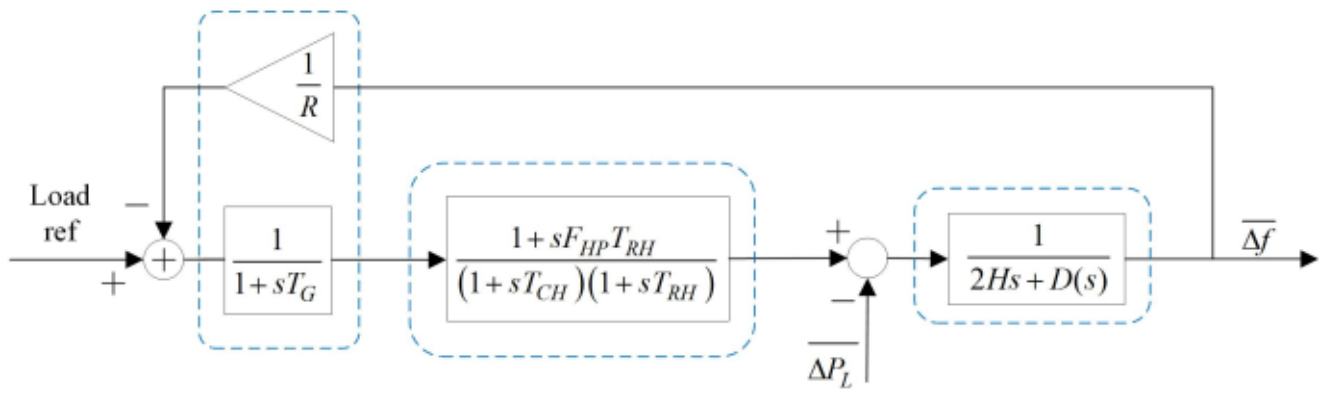

Fig. 3. The augmented block diagram of the power system

## III. Case Study

This section compares the frequency responses of the conventional and augmented LFC models, investigates the impact of dynamic loads on system performance, and analyzes the effects of IM parameters on system behavior.

To compare the conventional LFC model with the proposed augmented model, a typical power system dataset is adopted based on the parameters reported in Table I, using the references [18], [21]. A step change of 0.05 pu in the frequency-independent load (i.e. $\overline{\Delta P_L}$ ) is then applied and the corresponding frequency responses are simulated for 15 seconds. In this context, Fig. 4(a)–(d) illustrate the frequency responses of the two LFC models for system inertia constants of 5 s, 1 s, and 0.3 s, respectively. It is noteworthy that, in these figures, it is assumed that 50% of the total load consists of IM loads (i.e. $\alpha_{IM} = 0.5$ ).

Table I. The typical power system data used for simulation.

| Parameter | Value | Parameter | Value | Parameter | Value |
|---|---|---|---|---|---|
| $R$ | 0.05 | $F_{HP}$ | 0.3 sec | $k_{IM,n}$ | 0.8 |
| $T_G$ | 0.2 sec | $T_{CH}$ | 0.3 sec | $k_{IM,r}$ | 1 |
| $T_{RH}$ | 7 sec | $D, k_{ZIP}$ | 2 | $\tau_{IM}$ | 0.05 sec |

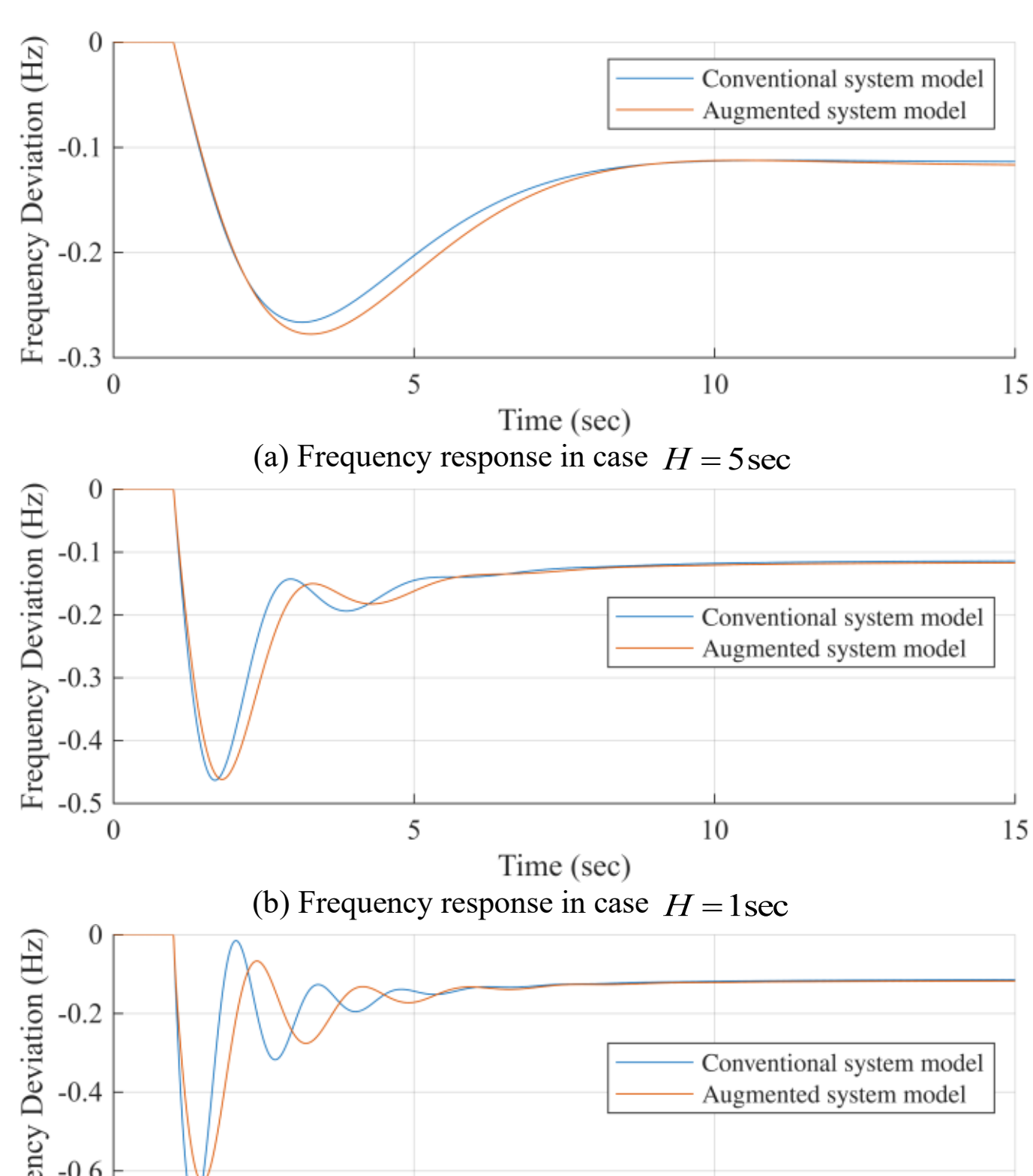

(a) Frequency response in case $H = 5$ sec

(b) Frequency response in case $H = 1$ sec

(c) Frequency response in case $H = 0.3$ sec

Fig. 4. Comparison between the conventional and enhanced system models.

As shown in Fig. 4, the frequency response of the proposed augmented model differs from that of the conventional system. The observed differences vary with the system inertia level. In Fig. 4(a), under high-inertia conditions, the augmented model exhibits a lower frequency nadir than the conventional model, indicating that the conventional model provides an optimistic estimation. In Fig. 4(b), when the system inertia is reduced to 1 s, both models show nearly identical nadir values; however, the augmented model demonstrates better RoCoF performance, suggesting a pessimistic estimation by the conventional model. In Fig. 4(c), under very low-inertia conditions, the augmented model outperforms the conventional model in terms of both frequency nadir and RoCoF, again indicating conservative predictions by the conventional model.

These results show that the conventional LFC model may yield significantly different frequency-response predictions under varying inertia conditions and therefore may not accurately represent system dynamics when dynamic load effects are considered.

To investigate the effect of the IM load proportion on the frequency response of the proposed model, four values of $\alpha_{IM}$ are considered: 0, 0.33, 0.66, and 1. The corresponding system frequency responses are illustrated in Fig. 5(a)–(c) for inertia values of 5 s, 1 s, and 0.3 s, respectively.

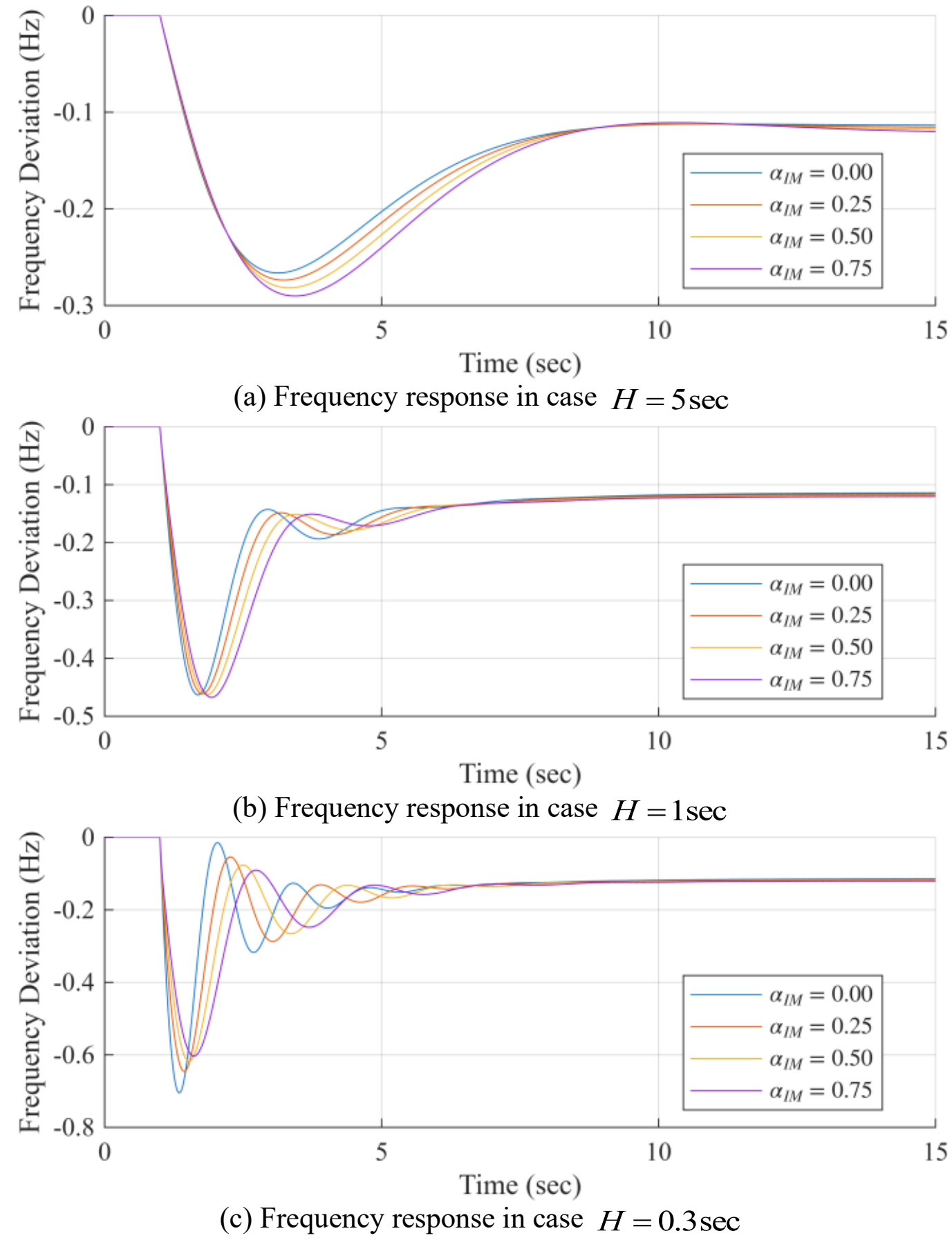

(a) Frequency response in case $H = 5\text{sec}$

(b) Frequency response in case $H = 1\text{sec}$

(c) Frequency response in case $H = 0.3\text{sec}$

Fig. 5. The effect of $\alpha_{IM}$ on the frequency response in different inertia levels.

According to these figures, under high inertia (5 s), the RoCoF remains nearly unchanged for all values of $\alpha_{IM}$, while the frequency nadir deteriorates as $\alpha_{IM}$ increases. For inertia of 1 s, the RoCoF improves with increasing $\alpha_{IM}$, whereas the nadir remains almost unchanged. For low inertia (0.3 s), both the RoCoF and the frequency nadir improve as $\alpha_{IM}$ increases. These results are also consistent with those in Fig. 4 and indicate that the FDL cannot be adequately represented by a simple damping factor, highlighting the need to explicitly consider dynamic load effects in system modeling.

Next, the influence of different FDL parameters on the frequency response is analyzed. This includes the parameters $k_{IM,n}$, $k_{IM,r}$, $\tau_{IM}$, and $k_{ZIP}$. In this regard, Figures 6(a)–6(d) investigate the effects of parameters $k_{IM,n}$, $k_{IM,r}$, $\tau_{IM}$, and $k_{ZIP}$, respectively. In each figure, four different values are considered for the corresponding parameter, resulting in four frequency response curves that allow observation of their individual impacts. These parameter values are indicated in the legends of each figure. It should be noted that the system inertia is assumed to be 0.5 s in all cases.

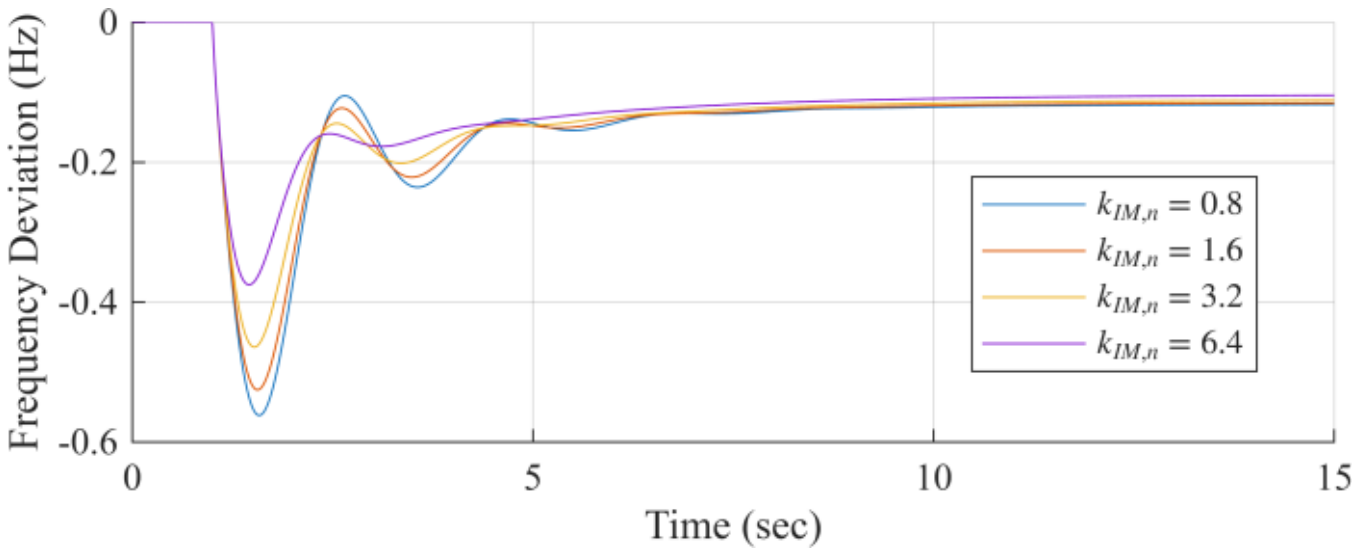

(a) Frequency response for different values of $k_{IM,n}$

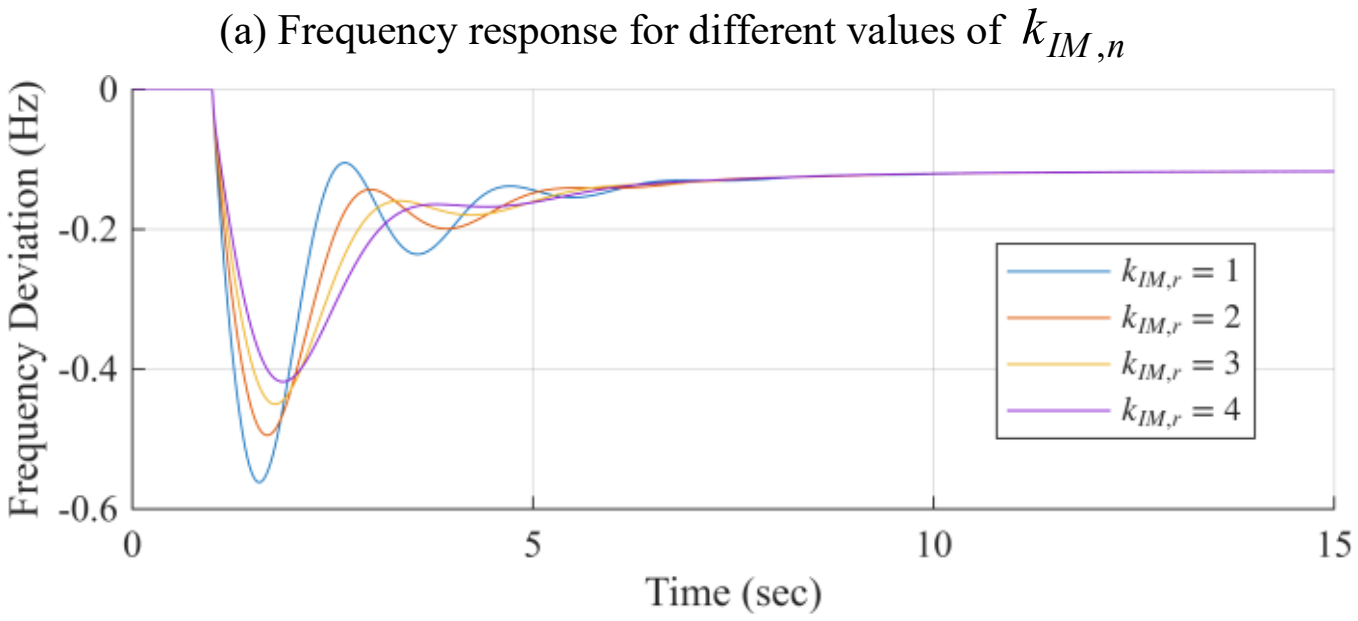

(b) Frequency response for different values of $k_{IM,r}$

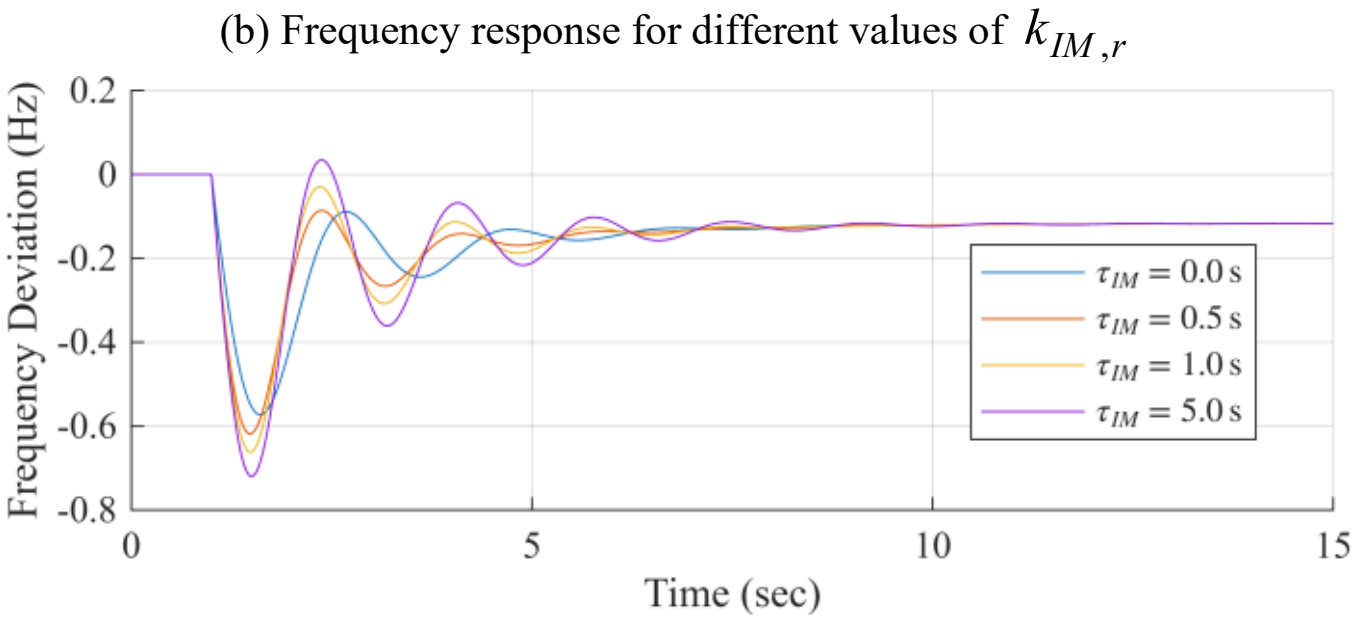

(c) Frequency response for different values of $\tau_{IM}$

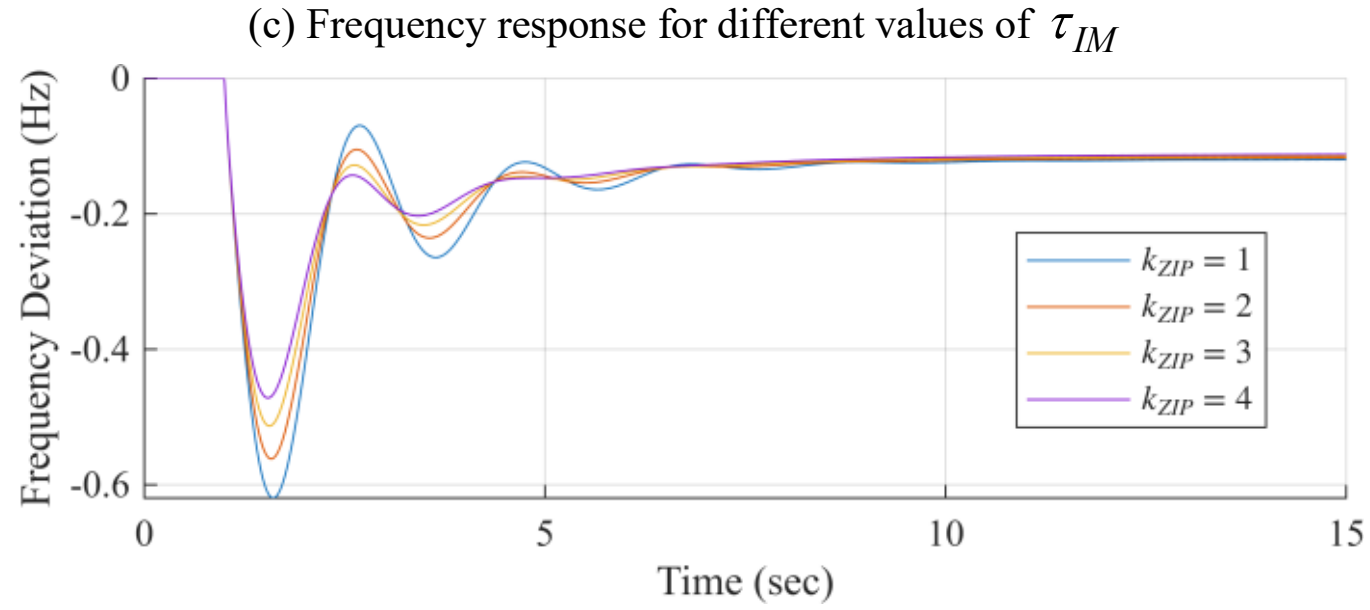

(d) Frequency response for different values of $k_{ZIP}$

Fig. 6. Impact of FDL parameters on the frequency response.

According to the obtained results, increasing either parameter $k_{IM,n}$ or $k_{IM,r}$ improves the frequency response in terms of both the nadir point and the RoCoF. However, the impact of parameter $k_{IM,n}$ is more pronounced on the nadir point, with a relatively smaller effect on RoCoF, whereas parameter $k_{IM,r}$ primarily influences RoCoF and has a comparatively weaker effect on the nadir. In contrast, increasing parameter $\tau_{IM}$, which represents the delay in load response, has an adverse effect on the frequency response, leading to deterioration in both the nadir point and RoCoF.

Additionally, it can be observed from Fig. 6. that increasing parameters $k_{IM,n}$, $k_{IM,r}$, and $k_{ZIP}$ generally improves the settling time of the frequency response, whereas increasing parameter $\tau_{IM}$ negatively affects it. The settling time refers to the time required for the frequency response signal to reach and remain within an acceptable range around its steady-state value [4]. A summary of these effects is presented in Table II. In this table, "↗" indicates improvement, while "↘" denotes degradation.

Hence, incorporating the IM model into the block diagram may alter the overall system behavior in different ways depending on the specific values of the IM load parameters. This demonstrates that the dynamic characteristics of the IM load can significantly influence the frequency response behavior observed in the overall power system LFC model.

Table II. Summary of FDL Parameter Effects on Frequency Response.

| Experiment | Result trend in | | |
|---|---|---|---|
| | RoCoF | Nadir point | *ST* |
| Increase in $k_{IM,n}$ | ↗ | ↗ | ↗ |
| Increase in $k_{IM,r}$ | ↗ | ↗ | ↗ |
| Increase in $\tau_{IM}$ | ↘ | ↘ | ↘ |
| Increase in $k_{ZIP}$ | ↗ | ↗ | ↗ |

## IV. Conclusion

This paper proposed an augmented LFC block diagram for improved frequency response analysis by incorporating both static and dynamic FDLs through considering induction motor dynamics. Unlike the conventional LFC model, which represents load effects using a single damping factor, the proposed approach replaces this simplified coefficient with a more general transfer-function. The results showed that the frequency response of the augmented model differs from that of the conventional model. Thus, conventional LFC models may not adequately capture system dynamics when dynamic load effects are considered. The parametric analysis further revealed that variations in induction motor parameters affect frequency-response characteristics differently, highlighting the importance of accurate dynamic load representation by taking the dynamic load model into account. Therefore, the proposed model provides a more representative framework for frequency response studies, control design, and operational planning in power systems.